\documentclass[graybox]{svmult}

\usepackage{mathptmx}       
\usepackage{helvet}         
\usepackage{courier}        
\usepackage{type1cm}        
\usepackage{makeidx}         
\usepackage{graphicx}        
\usepackage{multicol}        
\usepackage[bottom]{footmisc}

\makeindex             

\begin{document}

\title*{The ART of Cosmological Simulations }
\author{Stefan Gottl\"ober and Anatoly Klypin}
\institute{Stefan Gottl\"ober \at 
Astrophysical Institute Potsdam, An der Sternwarte 16, 14482 Potsdam, Germany
\email{sgottloeber@aip.de}
\and Anatoly Klypin \at Astronomy Department, New Mexico State University,
MSC 4500, P.O.Box 30001, Las Cruces, NM, 880003-8001, USA 
\email{aklypin@nmsu.edu}}
%
%
\maketitle

\abstract*{ We describe the basic ideas of MPI parallelization of the
  N-body Adaptive Refinement Tree (ART) code.  The code uses
  self-adaptive domain decomposition where boundaries of the domains
  (parallelepipeds) constantly move -- with many degrees of freedom --
  in the search of the minimum of CPU time. The actual CPU time spent
  by each MPI task on  previous time-step is used to adjust
  boundaries for the next time-step. For a typical decomposition of
  $5^3$ domains, the number of possible changes in boundaries is
  $3^{84}\approx 10^{40}$. We describe two algorithms of finding
  minimum of CPU time for configurations with a large number of
  domains.  Each MPI task in our code solves the N-body problem where
  the large-scale distribution of matter outside of the boundaries of
  a domain is represented by relatively few temporary large particles
  created by other domains. At the beginning of a zero-level
  time-step, domains create and exchange large particles. Then each
  domain advances all its particles for many small time-steps. At the
  end of the large step, the domains decide where to place new
  boundaries and re-distribute particles. The scheme requires little
  communications between processors and is very efficient for large
  cosmological simulations.}

\abstract{ We describe the basic ideas of MPI parallelization of the
  N-body Adaptive Refinement Tree (ART) code.  The code uses
  self-adaptive domain decomposition where boundaries of the domains
  (parallelepipeds) constantly move -- with many degrees of freedom --
  in the search of the minimum of CPU time.  The actual CPU time spent
  by each MPI task on a previous time-step is used to adjust
  boundaries for the next time-step.  For a typical decomposition of
  $5^3$ domains, the number of possible changes in boundaries is
  $3^{84}\approx 10^{40}$. We describe two algorithms of finding
  minimum of CPU time for configurations with a large number of
  domains.  Each MPI task in our code solves the N-body problem where
  the large-scale distribution of matter outside of the boundaries of
  a domain is represented by relatively few temporary large particles
  created by other domains. At the beginning of a zero-level
  time-step, domains create and exchange large particles. Then each
  domain advances all its particles for many small time-steps. At the
  end of the large step, the domains decide where to place new
  boundaries and re-distribute particles. The scheme requires little
  communications between processors and is very efficient for large
  cosmological simulations.}

\section{Introduction}
\label{sec:intro}

During the last 10 years new extensive observations of the Universe
were made using both ground-based telescopes and space instruments.
These measurements have provided new insights into the structure of
the Universe on various scales. A wide range of the electromagnetic
spectrum emitted by cosmic objects has been studied. The wavelengths
extend from very long radio wavelengths to energetic gamma rays.
This observational progress has been accompanied by considerable
effort in our theoretical understanding of the formation of different
components of the observed structure of the Universe: galaxies and
their satellites, clusters of galaxies, and super-clusters. A
substantial part of this theoretical progress is due to the 
improvment of  numerical methods and models, which mimic 
structure formation on different scales using a new generation of
massive parallel supercomputers. 

The collective effort of observers and theorists brought into being
the standard cosmological model, which is based on the idea that some
kind of dark energy contributes about 70~\% of the total
energy-density of the spatially flat Universe. The simplest form of
the dark energy is the cosmological constant, which was introduced in
1917 by Albert Einstein in his paper about the cosmological solutions
of the field equations of general relativity. The remaining 30\% of
energy density consists of matter. About 85\% of this matter is made
of unknown dark matter particles, which interact only
gravitationally. Only the remaining 15\% is the contribution of
``normal'' (baryonic)  particles, well known to particle physicists. This means
that at present we know the nature of only 5\% of the total energy in the
universe, the remaining 95\% is not yet understood.

The main process responsible for the formation of observed structures
is the gravitational instability. The initial seeds, which eventually
became galaxies and super-clusters and all the other structures, came
from the quantum fluctuations generated during the early inflationary
phase.  The power spectrum of these primordial fluctuations has been
confirmed by measurements of the temperature fluctuations of the
cosmic microwave background radiation.  These temperature fluctuations
tell us  the magnitude of the small density fluctuations in the
Universe when it was about 300~000 years old. One of
the key features of the standard model is its simplicity. The
expansion rate and the clustering properties are described by only few
parameters which are measured at present already with quite high
accuracy.

Since about 85\% of the matter consists of only gravitationally
interacting particles this dark matter forms the backbone structure
for all objects in the Universe from clusters of galaxies to dwarf
satellite galaxies. Baryonic matter  falls into the potential
wells created by the dark matter and forms luminous
objects. The nonlinear evolution of cosmological fluctuations can be
studied only numerically. The details of galaxy formation must be
followed using hydrodynamic simulations.  However, many features can
already be studied by semi-analytical methods which are based on the
evolution of dark matter halos as measured in the dark matter
simulations. Thus, numerical simulations are an important tool to
understand the formation of structure in the Universe.

The requirements for modern cosmological simulations are extreme: a
very large dynamical range for force resolution and many millions of
particles are needed. These requirements are just a reflection of the
vast range of masses and spatial scales in real astronomical
objects. For example, from dwarf galaxies to galaxy clusters the mass
spans about 7 orders of magnitude. The range of scales is also
enormous: from the inner structure of galaxies (sub-kiloparsec scales)
to cosmological distances of hundreds of megaparsecs.

\section{The Adaptive Refinement Tree (ART) code}
\label{sec:ART}

\subsection{History}
\label{subsec:hist}

To follow the evolution of the dark matter  in the expanding Universe
one has to solve the coupled system of the Poisson and Vlasov
equations.  These are the standard equations of motion for particles
interacting via gravity. Since the number of particles representing
the distribution of dark matter is very large, direct integration
of these equations is not possible. Over the last three
decades several methods have been developed to solve simultaneously
the Poisson equation for the gravitational potential and Newton's
equation for the acceleration for large numbers of particles (see
\cite{hockney:1988} for an overview). 

The Adaptive Refinement Tree (ART) code was build by a number of
people starting 1979. In its first version it was a particle-mesh code
written by A. Klypin in collaboration with A. Doroshkevich and
S. Shandarin (then at the Institute of Applied Mathematics in
Moscow). At that time the code used a cubic mesh to assign density and
to solve the Poisson equation. The Cloud-In-Cell algorithm was used to
find the density. Due to the limited computer resources in the early
eighties the first version of the code could handle only $32^3$
particles.

In 1995 A. Khokhlov \cite{khokhlov:1998} developed the Fully Threaded
Tree algorithm for Adaptive Mesh Refinement. He provided routines to
handle data structures in the new Adaptive Mesh Refinement (AMR)
scheme. Using the previous codes and Khokhlov's new algorithm, in
1996/97 A. Kravtsov \cite{kravtsov:1997,kravtsov:thesis} wrote the
first version of the ART code. This version of the code used the
OpenMP parallelization.

Because the parallelization of ART with OpenMP is not very efficient,
there was a need to substantially increase the scalability of the code
in order to use it on massively parallel computers. Starting 2000, we
developed MPI versions of the code based on the OpenMP code. The first
hybrid MPI+OpenMP code was written to simulate the evolution of 13
galaxy clusters using 8 nodes of the Hitachi supercomputer at LRZ. The
code was run as a farm of non-communicating OpenMP jobs.  On each of
the nodes the OpenMP parallelized code was running on 8 CPUs.

At that time the code also was modified to treat particles of
different masses and to have high resolution in some specific regions
of the computational box.  For example, in one of the simulations we
selected high density regions within a $80 {h^{-1}{\rm Mpc}}$ 
box and covered the
regions with many small mass particles, whereas the large scale tidal
field was represented by massive particles in the rest of the
computational box. To avoid numerical problems between the high- and
low- resolution regions, several layers with particles of increasing
mass were added.  Different high mass resolution areas were given
different MPI tasks.  The load balance in this case is not very
good because the evolution proceeds differently in different
simulations.  However, this is not a big problem because different
regions of the simulations could be in different stages of evolution
and we can use fewer MPI tasks once some of the jobs are finished.

In 2002-2003 we developed a full MPI+OpenMP code. The motivation to
have a hybrid MPI+OpenMP code is to address two issues: (a) The OpenMP
parallelization is not very efficient for large number of processors
and the code scaling depends on particular computer architecture. As
of 2007, the code scales well up to 4 processors on shared memory
computers such as Altix or SP5. For example, on a quad Opteron systems
the speedup is 1.8 for two processors and 2.4 for four processors. The
main bottleneck for OpenMP is the data locality. Thus, MPI is
necessary, if we want to use more processors.
(b) Significant memory is required by the code, and  OpenMP provides
the way to access a larger memory: when we use OpenMP, all the memory of a
node is accessible for the code. Therefore, we typically  use 2--4
processors per MPI task depending  on the memory requirements of
our computational problem and on the computer architecture.

Since 2004 the MPI version of the code was used on different computers
like the Hitachi and Altix of the Leibnizrechenzentrum Munich, the
Altix of the NASA Ames, the Opteron cluster at AIP, the p690 cluster at
Julich and the MareNostrum computer in Barcelona. Depending on the
requirements of our tasks and the computer architecture, we used 32 to
510 processors for our simulations.

Moore's law is roughly reflected also in the evolution of the ART
code: It's very first version could handle $32^3$ particles, whereas
20 years later the MPI version handles  $1024^3$ particles, a factor of
$2^{15}$ increase. Moore's law predicts a factor of $2^{13}$.

At present ART is a family of codes, which sprouted up from the same
AMR code written by A. Kravtsov \cite{kravtsov:1997,kravtsov:thesis}:
\begin{itemize}

\item OpenMP-only N-body code. This has been often used for simulations of
  isolated stellar dynamical systems \cite{Valenzuela2003,Colin2004}, for 
  a computational box with up to $256^3$ particles \cite{Colin2006}, or for 
  a single high-resolution region in a large computational box 
  \cite{gottloeber:2003}.
\item MPI+OpenMP N-body code. This code is used for large cosmological
  simulations.
\item OpenMP N-body+hydro code. Examples of using the code include
  simulations of clusters of galaxies \cite{Kravtsov2005,Nagai2007},
  large-scale distribution of gas in the Local Supercluster \cite{Kravtsov2002}, and formation of galaxies \cite{Ceverino2007}.
\item MPI N-body+hydro code. This was written by D.~Rudd and
  A.Kravtsov \cite{Rudd2008}.  N.~Gnedin incorporated radiative transfer code into the ART  hydro code \cite{Gnedin2008}.
\end{itemize}
Here we discuss parallelization of the MPI+OpenMP N-body code.  
\subsection{Method}
\label{subsec:method}

The code starts with a regular cubic grid, which covers the entire
computational volume and defines the minimum resolution of the
simulation.  If the mass in a cell exceeds some threshold, the cell
can be split into eight cells each half the size. If mass in any of
the new cells is still above the threshold, the cell can be split again. 
In order to avoid too large jumps in the sizes of adjacent
cells, the code enforces splitting of cells in such a way that
refinement levels of any adjacent cells differ by not more than one
level. In other words, if $n$ is the level of a cell, than its
immediate neighbor may be only a cell on levels $n-1,n,n+1$.  The code
constructs meshes of arbitrary shape covering equally well both
elongated structures (such as filaments and walls) and roughly
spherical dark matter halos.  The meshes are modified to adjust to the
evolving particle distribution by creating new cells and by destroying
old ones. The threshold for the refinement is a free parameter, which
is typically 2-4 particles in a cell. The algorithm of the refinement
is very flexible and can be easily adjusted for a particular
problem. For example, we can allow the construction of the refinements
only in some specified area of the computational volume. This is done
by constructing a map of refinements: only cells marked for refinement
are allowed to be split.

The ART code integrates trajectories of collisionless particles by
employing standard particle-mesh techniques to compute particle
accelerations and advance their coordinates and velocities in time
using the leap-frog scheme.  The time-step in ART code depends on the
resolution level: the higher the level (and the density), the
smaller the time-step. The time-step decreases by a factor of two
with each refinement level.  In cosmological simulations the
refinement can reach 10 levels, which gives 1024 times smaller time-steps 
as compared with the zero-level time-step.  Typically a
cosmological simulation has between 300 and 1000 zero-level time-steps
or even more in the case of very high resolution runs. The ART code
should run with sufficiently small time-step so that the maximum
displacement does not exceed a fraction of a cell. Typically, it
should be below 0.20-0.25. This corresponds to the rms displacement in
the range 0.05-0.1. If the maximum displacement goes above unity, the
code may become unstable and it should be restarted from the very
beginning with a smaller time-step.

To solve the Poisson equation the code uses the FFT solver at the
zero-level of refinement and a multilevel relaxation method with
odd-even Successive Over-relaxation with Chebyshev acceleration at
each non-zero level.

\subsection{The MPI version of the code}
\label{subsec:MPI}

The basic idea of MPI parallelization of the ART code is to decompose
the simulation volume into rectangular  domains.  Each
MPI task handles one domain and employs the OpenMP version of the
code.  Communications
between MPI tasks occur only at the beginning of each zero-level
time-step. Each MPI task receives information about the mass
distribution and velocity field in the whole computational
volume. This information is accurate enough to advance particles
handled by the task to the next time-step.  The information comes in
the form of massive  particles, which represent mass distribution and
velocity away from the domain of the MPI task. At the end of the
zero-level step these additional particles are discarded and the whole
process starts again.

\paragraph{\bf Domain decomposition} 

\begin{figure}[t]
\sidecaption[t]
\includegraphics[height=0.6\textwidth]{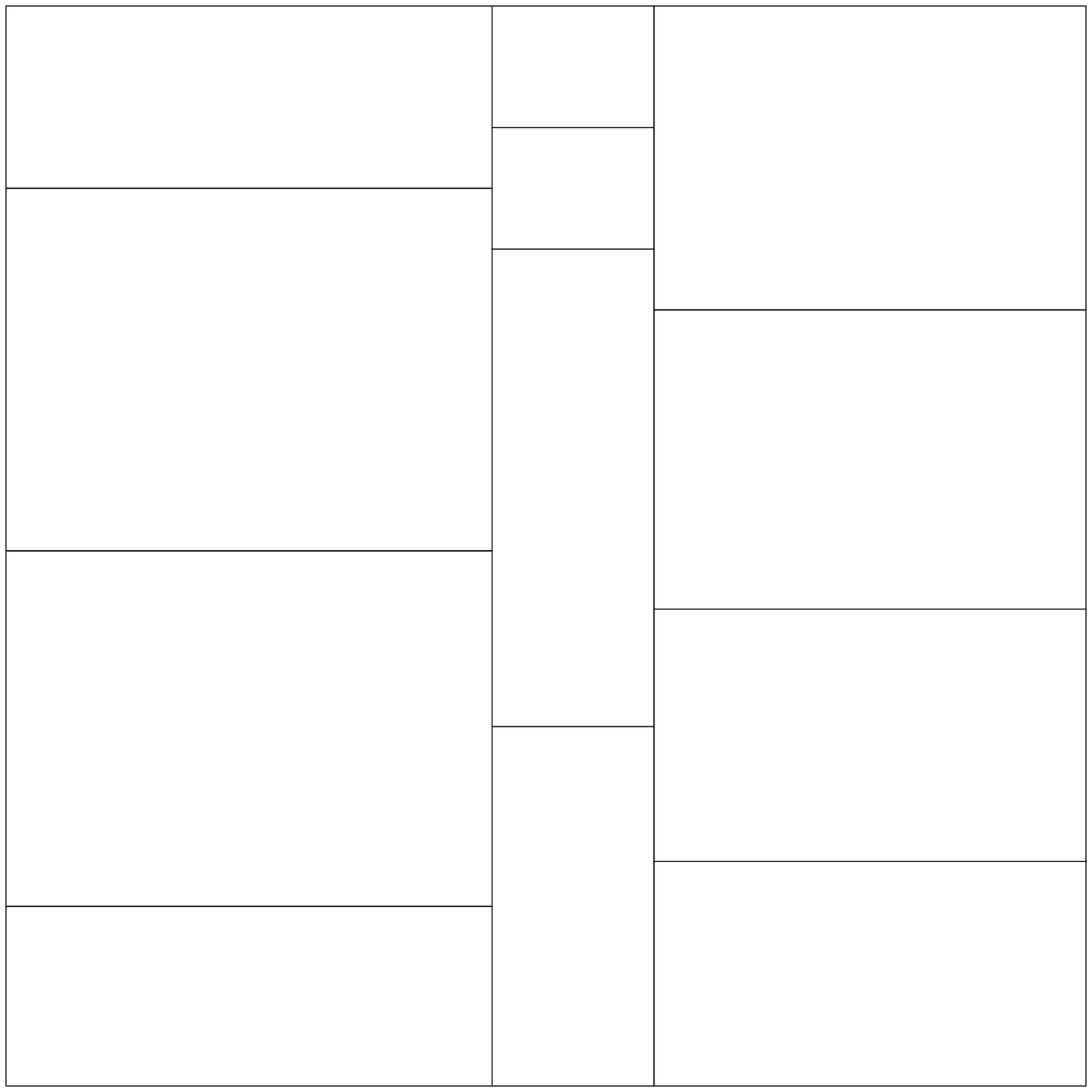}
\includegraphics[height=0.45\textwidth]{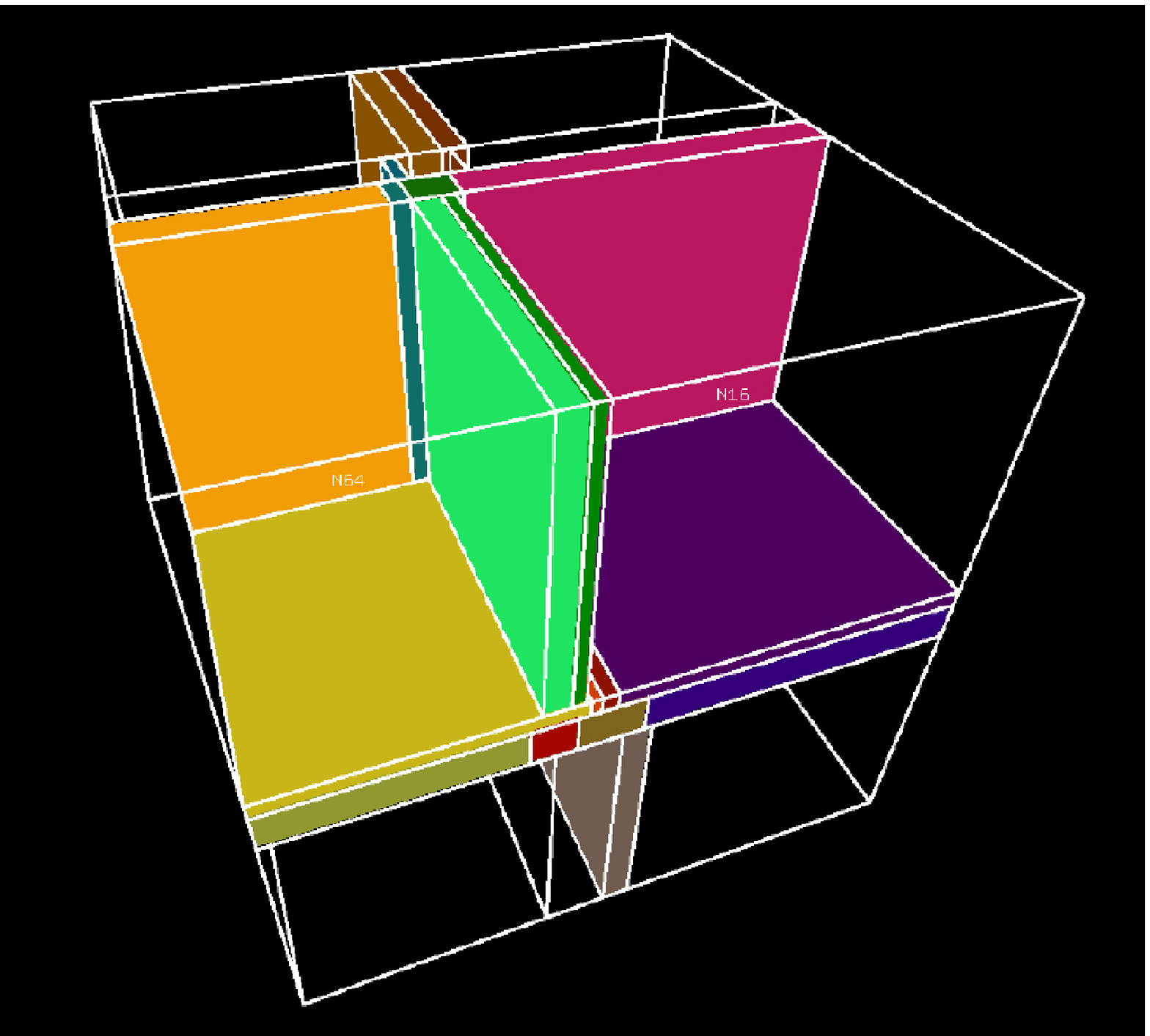}
\caption{ {\it Left:} Example of the 3x4 domain decomposition in 2D.
  Boundaries of domains are aligned in x-direction, but there is no
  alignment of domains in y-direction. The total number of degrees of
  freedom is 11. Each boundary can be adjusted in order to minimize
  the maximum CPU of MPI tasks.  {\it Right:} Domain decomposition in
  3D for a high resolution simulation, which was dominated by a single
  halo in the center of the box. The code was configured to have 4x4x4
  domains. This is an extreme case with some domains being very
  elongated. Typical simulations have more even distribution of
  domains. There is no problem in solving the force of gravity even in
  these extreme conditions: this is what AMR-type codes are
  designed for. Yet, the code is not efficient because most of the CPU
  time goes to the central region and there is little left for the rest
  of the domains. }
\label{fig:domain}
\end{figure}

We use rectangular domains for MPI parallelization. The whole
simulation volume - a cube - is split into non-overlapping fully
covering parallelepipeds.  The boundaries of the parallelepipeds can
move as time goes on in order to equalize the load of different MPI
tasks. In Fig. \ref{fig:domain}, left panel we show an example of a possible
splitting of the computational volume in the two-dimensional case.
Note, that boundaries in x-direction are aligned, but they are not
aligned in y- (and in 3D in z-) directions. 

Each boundary can move only at the beginning of a zero-level
time-step. Once the zero-level time-step is completed, information on
CPU time consumed by different MPI tasks is used to adjust the
boundaries to improve the load balance.  At present boundaries of the
domains can have only discrete positions: they can only be placed at
boundaries of the zero-level mesh.  The number of degrees of freedom
to move domains can be very large. It depends on the number and the
configuration of the domains. For the example in
Fig. \ref{fig:domain}, left panel there are 3 domains in x- and 4 domains in
y- direction. In the general case of the
division of the volume by $n_x$ domains in the x-direction and $n_y$,
$n_z$ in y- and z-directions, the number of degrees of freedom is 
$(n_x-1)[(n_y-1)(n_z-1)+(n_y-1)+1]$.

There are different ways of using this large number of degrees of
freedom to improve the load balance. The current version of ART
provides two routines for load balance.  The first routine assumes
that within each domain $(i,j,k)$ the CPU time $T(i,j,k)$ is
homogeneously distributed. In other words, the density of the CPU time
is treated as a piece-wise constant function. Each boundary can have only
three positions: current, current plus one zero-level cell, and
current minus one zero-level cell. The code loops through a very large
set of  configurations of boundaries (up to 60000) and
finds the minimum value of the maximum expected CPU time in the
domains. The minimization routine is very fast - it takes only a
fraction of a second to find the minimum. By design, it minimizes the
maximum CPU time of a MPI task. This works reasonably well when the
system is evolving slowly and the maximum is not jumping from one area
to another. The constraint that boundaries can move only by one cell
works well for systems, which evolve slowly and for which the load
balance is already reasonable. In this case the code tunes the load
balance.

For early stages of evolution and for quickly evolving systems ART
uses a second algorithm of load balancing. The algorithm is to
equalize the load balance along each direction. It starts with
the x-direction. All CPU times are summed up for all tasks, which have the
same x-boundaries. In Fig. \ref{fig:domain}, right panel this gives three
numbers each being a sum of CPU time of domains in the y-direction with
the same x-boundaries. We can describe each domain by a
triplet of integers $(i,j,k)$, where integers are in the ranges
$[1,n_x]$, $[1,n_y]$, $[1,n_z]$. The procedure of summing up CPU times
gives $\sum_{j,k} T(i,j,k)$. Assuming that the CPU time is constant
inside boundaries of each domain, we get a piece-wise linear function of
CPU time from $x=0$ to given $x$. We can place new $x$-boundaries of
domains in such a way that each sum of domains with given
$x$-boundaries has the same CPU time: $\sum_{j,k} (i,j,k) =T(i)=
\rm{const}$. We then repeat this procedure for $y$- and $z$-directions.

The configuration of domains - how many in each direction and the
boundaries of the domains - is in a configuration file.  The
Fig. \ref{fig:domain}, right panel gives an example how the code can adjust
boundaries of domains in its effort to load-balance the run. In this
extreme example there was one large halo close to the center of the
computational box and few smaller halos and filaments around it.  The
code was using $4 \times 4 \times 4$ domains. After some period of
evolution, the code evolved in such a way that it had eight large
domains in the corners of the simulation box, which contained only a
small number of particles. Most of the computational effort was in the
smaller domains, which cover the central region of the box where the
massive halo and few smaller ones have formed.

\paragraph{\bf Exchange of information between MPI tasks}

At the beginning of each zero-level time-step the MPI tasks exchange
information. This is very infrequent. The main idea for the
information exchange is the same as in TREE codes: the mass
distribution at large distances can be approximated roughly when
forces are calculated. In the ART code this idea is implemented by
creating increasingly more massive particles with increasing distance
from the boundaries of a domain handled by an MPI task.  In addition,
every domain is surrounded by a buffer zone, from which it receives
primary (small) particles. Particles are not averaged in this buffer
zone.  The width of the buffer is a parameter. We typically use
(0.5-1) of the zero-level cell.

Thus, each MPI task has three types of particles: (1) primary
particles of low mass, (2) low mass particles in the buffer zone, and
(3) progressively more massive temporary particles. The set of all
particles in each domain covers the whole computational volume. Each
MPI task handles the whole volume and there is no other exchange of
information between MPI tasks until the end of the given zero-level
time-step. Only at the beginning of a zero-level time-step the
temporary particles are created and exchanged.  During one zero-level
time-step each MPI task advances all its particles (primary as well as
temporary).

Once the time-step is finished, the CPU time consumed by every MPI
task is gathered by the root task which decides how to move the
boundaries of the domains in order to improve the load balance. The
primary particles are re-distributed so that they reside on tasks,
which handle the domains. Then the process starts again: exchange
of buffer particles, creation and sending of  temporary particles.

Massive particles are created in the following way. Each domain (a
parallelepiped) is covered by a hierarchy of grids. The first grid has
cell size equal to the zero-level mesh. The second mesh has cells twice
the size of the zero level, the third mesh has cells twice the size of 
the second level, and so on for higher level meshes.  There are 4 levels
of meshes for construction of large temporary particles. We find mass,
average velocity, and center of mass of all primary particles in each
cell for each mesh.  This creates temporary massive particles,
which are sent from one domain to another to trace the external
gravity field. The level of grid, from which a temporary particle is
taken, depends on the distance to the boundary of the domain to which
the particle will be sent: the larger the distance the higher is the
level of the grid. If the zero-level mesh has $n_g$ cells along each
direction and $L$ is the length of the computational box, then the
zero-level cell has size $d_0=L/n_g$.  This length provides a scale
for the auxiliary meshes.  Within a shell of 8 zero-level cells, which
surrounds a given domain, the mesh used for creating temporary
particles is $d_0$. The next shell of 8 cells gives larger particles
taken from the second level mesh. The averaging size of the mesh is
$2d_0$.  For the third shell of 8 cells the averaging size is 
$4d_0$. Finally everything else is covered with $8d_0$ cells. (The
code is written for arbitrary number of mesh levels.)

We can estimate the number of temporary particles for each domain by
assuming that the distribution of mass is not too inhomogeneous. In
this case each domain has approximately $K_i=n_g/n_i$ zero-levels
cells in each $i-th$ direction.  Further assuming that the number of
domains in each direction is the same $n_x=n_y=n_z =n$, we estimate
the number of temporary particles $N_{\rm temp}$ on all levels:
\begin{equation} N_{\rm temp} =
\frac{n_g^3}{8^3}+7\sum_{j=1,3}\left(\frac{K+16j}{2^j} \right)^3-K^3,
\end{equation} 
\noindent where $K=n_g/n$ is the number of zero-level cells in 1D in
each domain. For typical values $n_g=256$, $n=4-6$, we get $N_{\rm
temp}=(2-4)\times 10^5$.  Most of the particles are coming from the
first (high resolution) shell.

In the same fashion we also can estimate the number of primary
particles in the buffer zone:
\begin{equation}
N_{\rm buffer} = \frac{N_{\rm part}}{N_{\rm domains}}\left[(1+\frac{2dx\cdot n}{n_g})^3-1\right],
\end{equation}
\noindent where $N_{\rm part}$ is the total number of the primary
particles in all domains, $N_{\rm domains}=n^3$ is the number of the
MPI tasks, and $dx$ is the width of the buffer zone in units of the
zero-level cell.  For typical numbers $N_{\rm part}=1024^3$, $N_{\rm
  domains}=6^3$, $dx=0.5$, and $n_g=256$, we get $N_{\rm buffer}
=3.5\times 10^5$. This should be compared with number of primary
particles of the MPI task $N_{\rm main} = \frac{N_{\rm part}}{N_{\rm
    domains}}\approx 5\times 10^6$. This means that the overhead of
the domain decomposition method is about 10 percent. The actual
overhead can be larger. For large cosmological simulations with sizes
of simulation boxes 100~Mpc and larger we actually measured overheads
close to the theoretical 10~percent, when the number of domains was
$\sim 5^3-6^3$ and the number of processors was 250-500.  Simulations
of small high-resolution regions embedded into a large computational
boxes get less efficient as the size of the high resolution region
gets closer to the size of a zero-level cell. Simulations of an
isolated halo, which is only a few zero-level cells across do not
scale and cannot be done efficiently with the percent version of the
code.  For the code to be efficient the rule of thumb is that the
number of large virialized systems consuming most of the CPU should be
larger then the number of MPI tasks. The problem with the scalability
of very large single-object simulation is not specific to the ART
code. To improve the scalability the decomposition should be done 
on high levels of the force refinement.

Using the estimates of the number of the particles, we can find how
much data should exchanged between MPI tasks. Each particle in the
simulation needs 6 numbers with double precision (8 bytes each) and
three auxiliary single precision variables. The total is 60 bytes per
particle.  Thus the total amount of data each MPI task receives and
sends is about 100Mb for the typical values presented above. 
Specifically, we use nine {\it mpi\_alltoallv} calls to distribute the
particles among the MPI tasks.

\paragraph{\bf Input - Output}
\label{subsec:IO}

Each of the $N_{\rm MPI}$ tasks reads its own files with information
about parameters, coordinates, velocities, particle masses and ids and
refinement levels of each particle. These files are stored in $N_{\rm
MPI}$ directories. When the code starts, every MPI task goes to its
subdirectory and reads its files. In the sense of structure, there is
no difference between snapshot files and the re-start files so that
one can easily restart from any earlier saved snapshot.  In each
subdirectory there are additional files that contain protocols of
running the job, in particular the CPU time spent at each time-step by
the given MPI task.

\begin{figure}[t]
\sidecaption[t]
\includegraphics[width=0.5\textwidth]{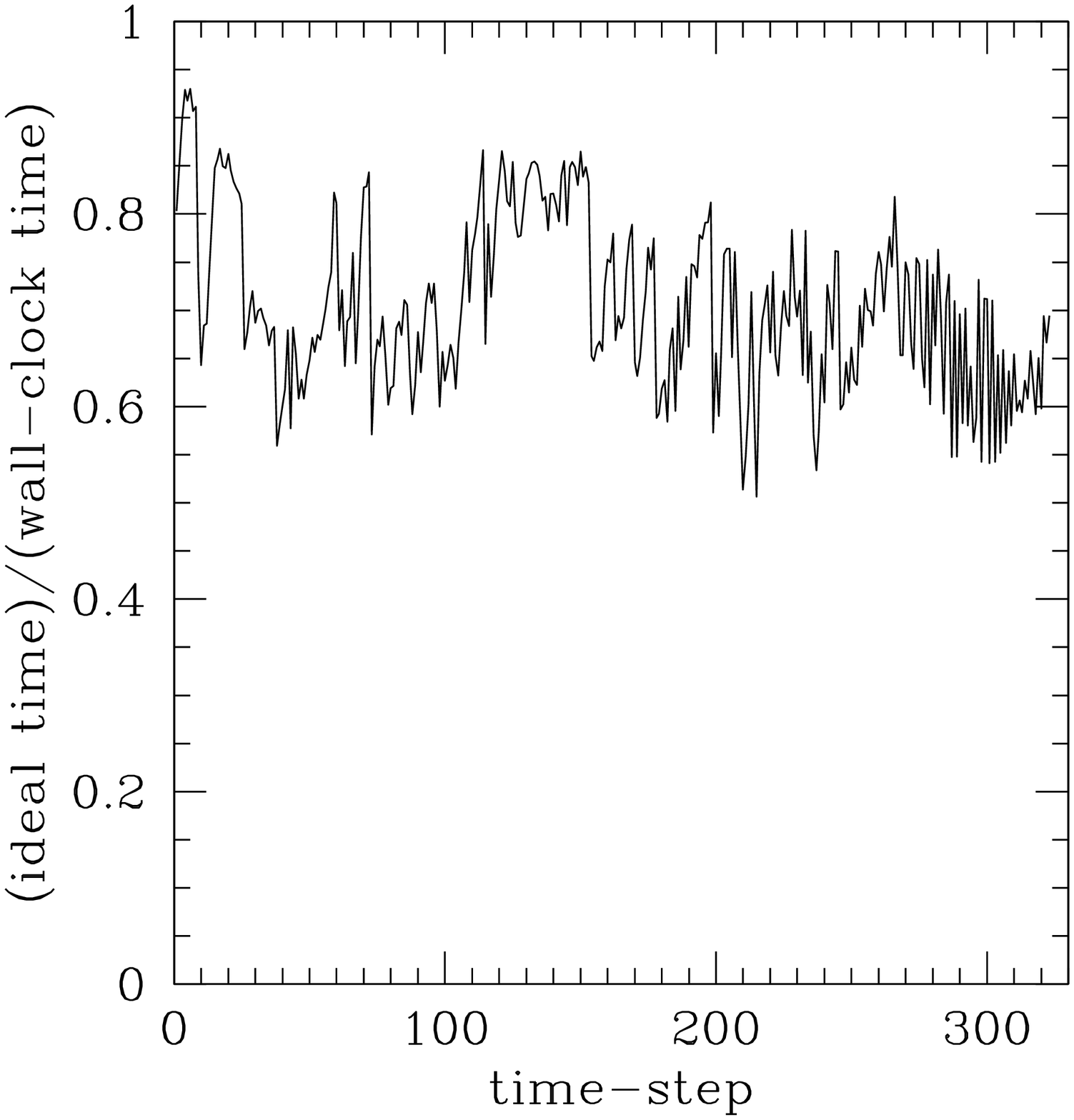}
\includegraphics[width=0.5\textwidth]{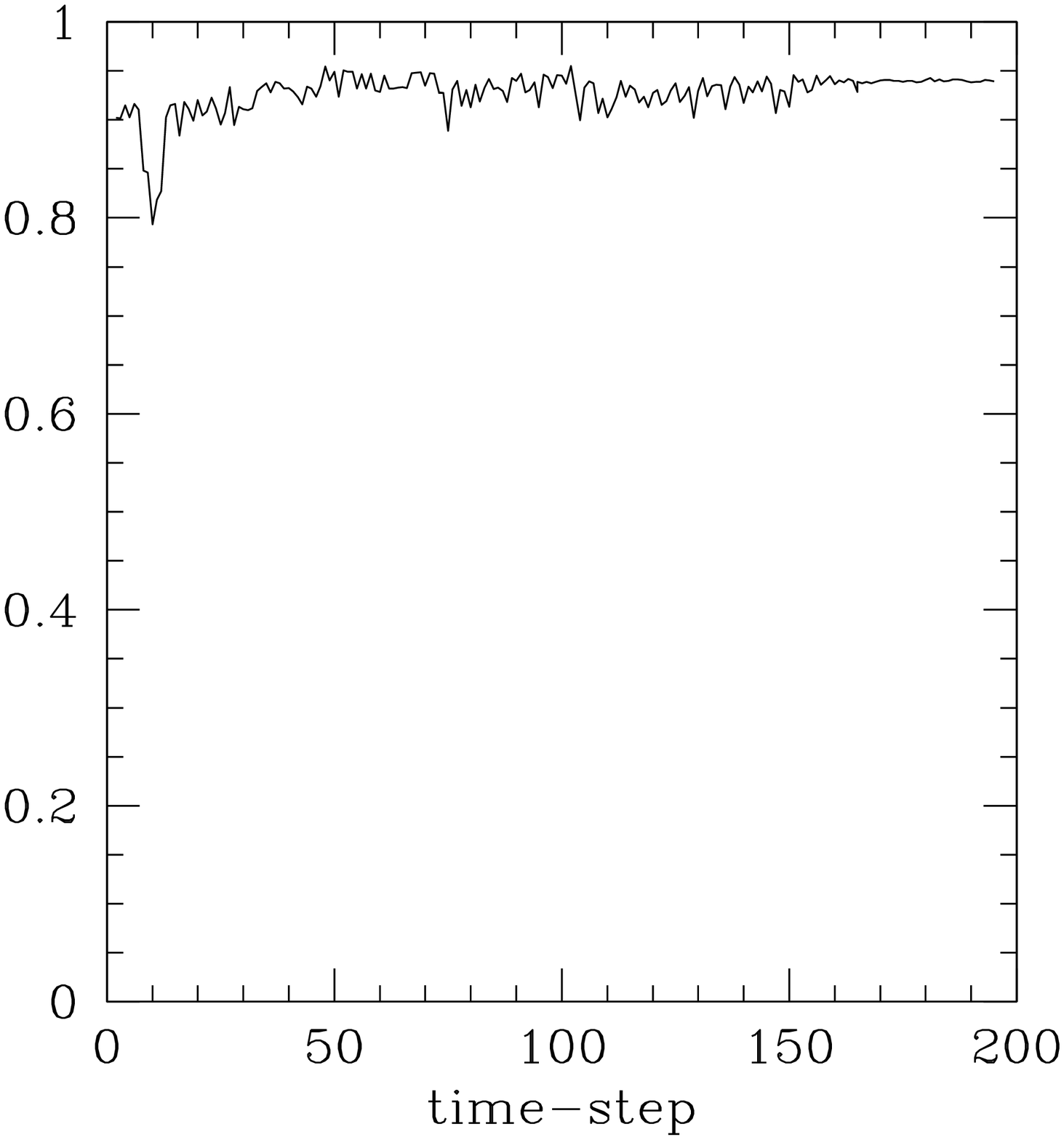}
\caption{ Load balance for two runs with $1024^{3}$ particles.   We show the ratio
  of the ideal CPU time to the wall-clock time used for each
  zero-level time-step.   The ideal time is calculated as the sum of CPU time
  over all procesors divided over the number of processors. The wall clock time
  is the time of the slowest MPI task. {\it
    Left:} Computational box is $160{h^{-1}{\rm Mpc}}$. In this case
  few large clusters dominate the evolution of the system and cause
  some load imbalance. The run was using 504 processors of the Altix system at
  LRZ Munich.   {\it
    Right:} Load balance for a run with $1024^{3}$ particles in a
  $1000{h^{-1}{\rm Mpc}}$ computational box.  The code used 500
  processors (125 MPI tasks with 4 processors in each task) of the
  Altix 4700 system at LRZ Munich and, starting with the time-step
  150, on Columbia Altix 3700 system at Nasa Ames.   Initial load imbalance
  (time-step about 20) occurs when the system starts to open
  refinement levels in different parts of the box. Once non-linear
  structures appear all over the computational volumes, the code
  adapts and equalizes the load reasonably well.}
\label{fig:loadbalance}
\end{figure}

The root task writes an additional file, which provides details of the
distribution of CPU time among different MPI tasks, it contains the
maximum, minimum, and average CPU time per zero-level time-step in
units of seconds as well as the CPU time used by each MPI task in
units of maximum CPU time. This tells us about the load balance. As an
example the load balance for a simulation of a computational box
of $160 {h^{-1}{\rm Mpc}}$ side length with $1024^3$ particles is shown in
Fig. \ref{fig:loadbalance}, left panel. For this simulation we used
252 nodes with 2 CPU per node on the Altix of LRZ Munich. 
Fig. \ref{fig:loadbalance}, right panel shows the load balance in
another simulation with $1024^3$ particles. Due to the lower number of MPI tasks
and the larger volume of the box ($1000 {h^{-1}{\rm Mpc}}$ side length) the
averaged volume which each MPI task has to handle is much
larger. Therefore, in this case the balance was significantly better.

\begin{figure}[t]
\sidecaption[t]
\includegraphics[width=\textwidth]{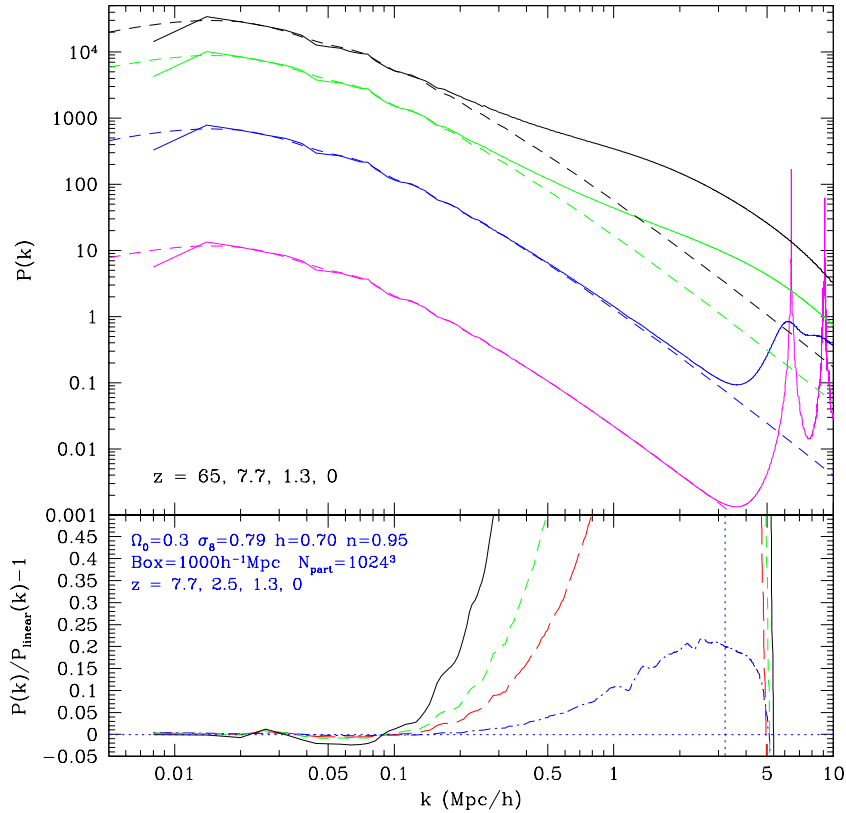}
\caption{Growth of perturbations in the $\Lambda$CDM model. The top
  panel shows the evolution of the power spectrum $P(k)$ in the
  simulation (full curves) as compared with the linear theory (dashed
  curves). From bottom to top the curves correspond to decreasing
  redshifts: the lowest curve is for the initial conditions ($z=65$)
  and the top curve is for $z=0$.   The deviation in the first
  harmonic (the smallest $k$) is due to small statistics of the longest
  waves. Two strong spikes at large $k$'s are above the Nyquist
  frequency: the N-body code does not ``see'' them.  Bottom panels
  show deviations from the predictions of the linear theory.
  Fluctuations on large scales (small $k$'s) grow according to the
  linear theory.  The dot-dashed ($z=7.7$) and the long-dashed curves
  ($z=2.5$) show that non-linear evolution increases the power
  spectrum on all scales proceeding from high $k$, where the
  non-linear effects are strongest, to low $k$, where the effects are
  weakest.  The vertical dotted line shows the Nyquist frequency of
  particles. Perturbations with the frequencies above the Nyquist
  frequency do not grow  in linear and in quasi-linear regimes.}
\label{fig:growth}
\end{figure}

Our scheme of MPI parallelization has one significant positive
feature: it has very little communications. Communications happen only
once every zero-level time-step, when the particles are re-arranged
between different MPI tasks.  For a typical simulation this happens
200-500 times during the whole run (so, every 30 min - 1hr of
wall-clock time for a large run). During that stage every MPI task
receives about 100-500Mb of data. Then there will be no communications
till the next time-step.  The scheme has its overheads and
limitations. CPU time required to handle particles in a narrow buffer
zone around each domain is a loss. Massive particles, which represent
external density field is also a loss, but the CPU required for them
is very small: a fraction of a percent of the total CPU. As long as the
number of particles in the buffer zone is small, the code works
reasonably well.  Thus, good load balance can be reached in large
cosmological runs that cover the whole computational box with
equal-mass particles. In case of multi-mass runs which resolve only a
certain region of the box (as shown in Fig. \ref{fig:domain}, right
panel) the load-balance is typically worse.

\section{Recent simulations run with the ART code}

In this section we present some results obtained from a series of
recent simulations done with the MPI version of the ART code at
different supercomputers.  In simulations with $1024^3$ particles we
identify 1-2 million halos. This is an excellent database for many 
different kinds of statistics. 

Figure~\ref{fig:growth} shows the evolution of the power spectrum of
perturbations in a large simulation of the $\Lambda$CDM model: 1Gpc
box with $1024^3$ particles. The longest waves in the simulation have
small amplitudes and must grow according to the linear theory. This
indeed is the case, as seen in the bottom panel. Note, that a small
dip $\sim -2\%$ at $k=0.05-0.07{h^{-1}{\rm Mpc}}$ is what the
quasi-linear theory of perturbations predicts \cite{Crocce2008}.  The
plot also shows the main tendency: in the non-linear stage the
perturbations at the beginning grow faster than the predictions of the
linear theory (we neglect a possible small negative growth extensively
discussed in \cite{Crocce2008}). At later stages the growth
slows down, which is seen as bending down of $P(k)$ at high
frequencies at $z=0$.

The evolution of the power spectrum at high frequencies (comparable
with the Nyquist frequency $k_{\rm Ny}$) is very challenging for
N-body codes. Note that the initial power spectrum matches the linear
theory nearly perfectly down to $k_{\rm Ny}$. This is done by
perturbing particles from a homogeneous grid. If initial conditions
were started with a random distribution, the initial spectrum would
have been dominated by the shot noise, whose amplitude would have been
$P_{\rm noise}(k)=1$: three orders of magnitude higher than the $P(k)$
at $k_{\rm Ny}$. There is a danger that the high-power fluctuations
above $k_{\rm Ny}$ (discreteness effects) may affect the growth of
real low-frequency waves. This does not happen as the comparison of
$P(k)$ at $z=65$ and $z=7.7$ shows. At $z=7.7$ the rms density
fluctuations are $\delta\rho/\rho\approx 0.4$ and the system is
approaching the non-linear stage at high frequencies. Yet, the gradual
upturn in the power spectrum seen in the low panel continues all the
way to $k\approx 0.8k_{\rm Ny}$. This suppression of the discreteness
effects is due to a carefully chosen force resolution. Initially we place
particles in every other resolution cell. As a  result, the code effectively
suppreses the discrerteness effects and does not impede the growth of
real fluctuations at $k <k_{\rm Ny}$. At later stages of evolution,
when most of small-scale fluctuations have grown and collapsed, we start
easing the refinement condition and gradually increase it to normal
2-4 particles per cell.

\begin{figure}[t]
\sidecaption[t]
\includegraphics[width=0.5\textwidth]{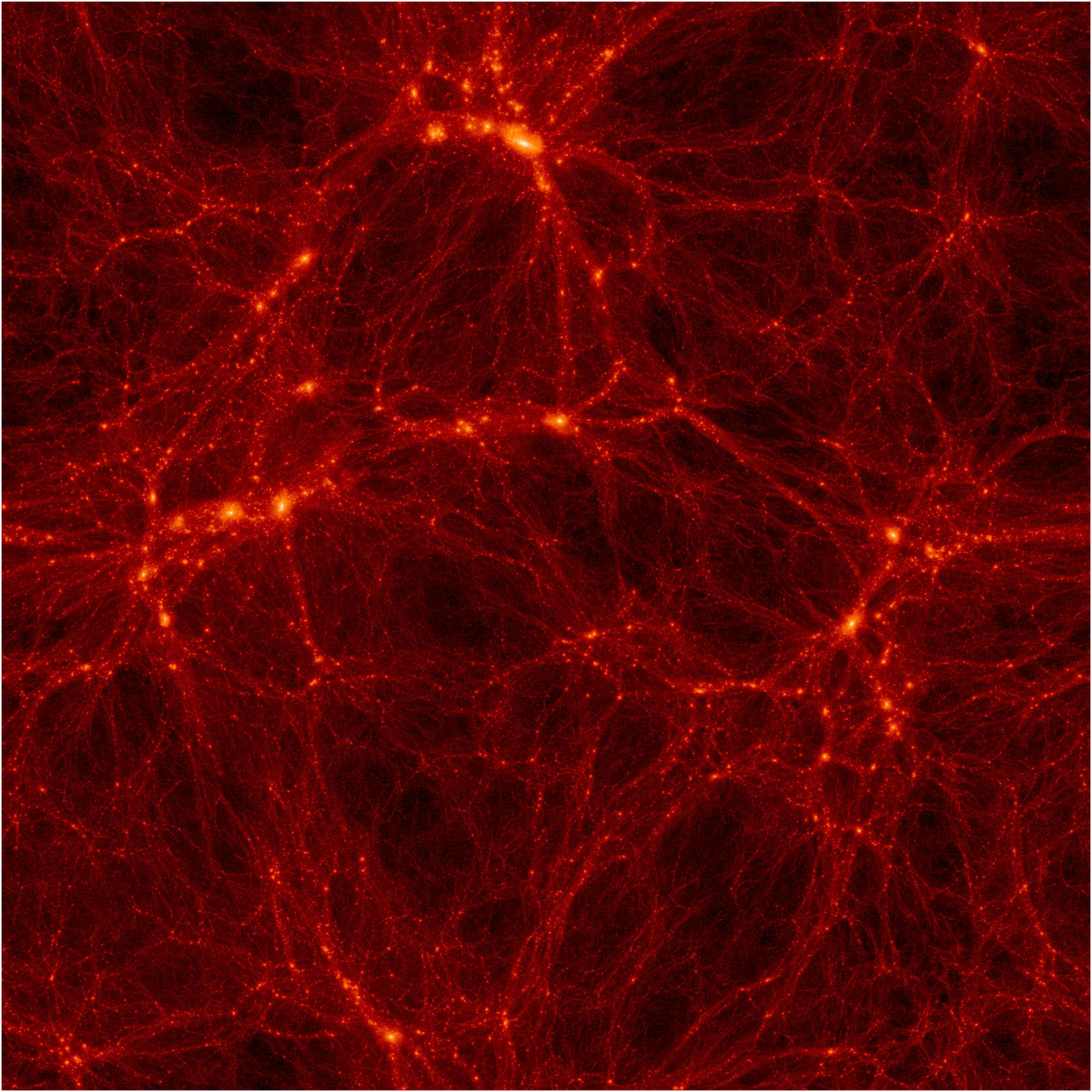}
\includegraphics[width=0.5\textwidth]{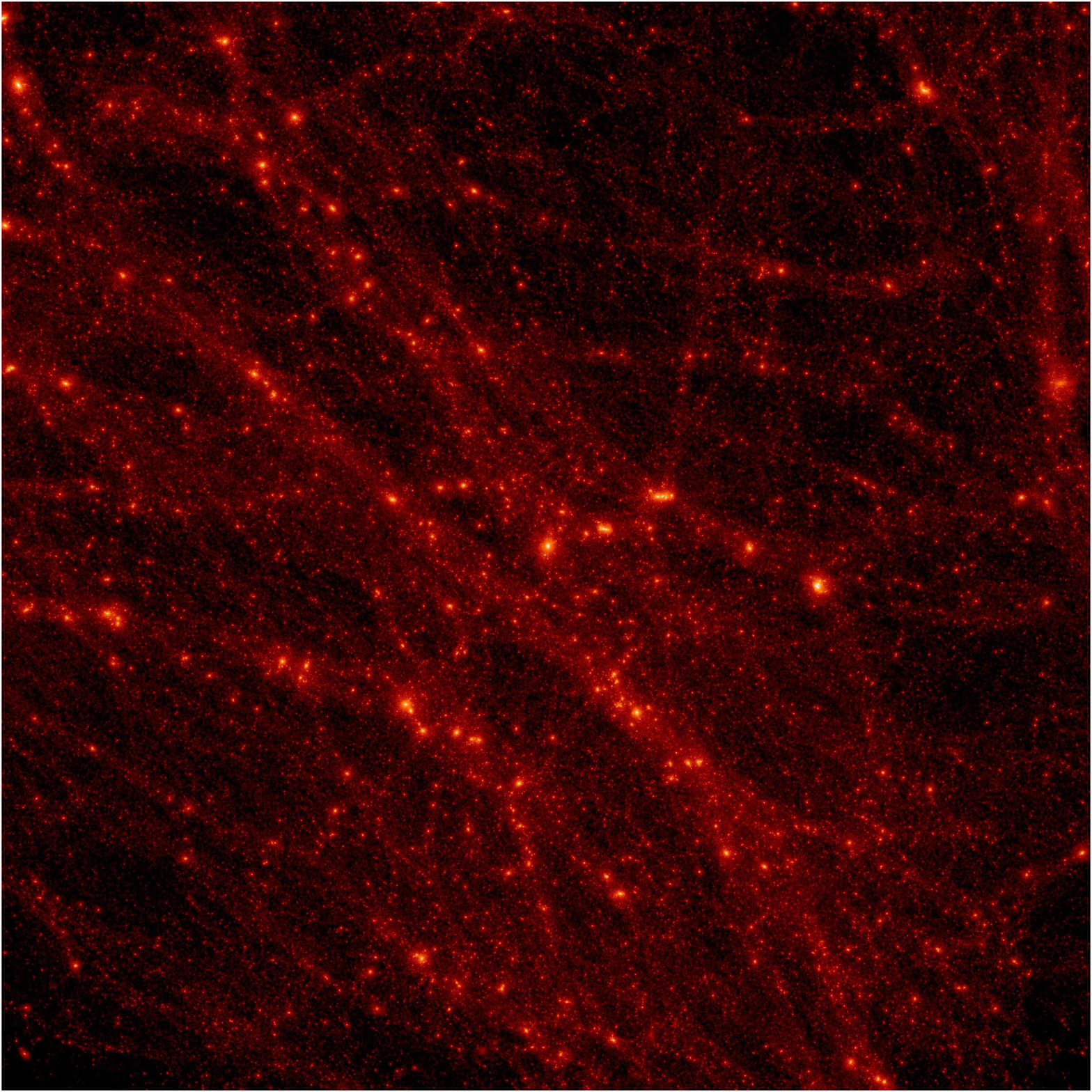}
\caption{{\it Left:} A slice through a $160 {h^{-1}{\rm Mpc}}$ box with $1024^3$
  particles. The color codes the dark matter surface density in this
  10~Mpc thick slice. {\it Right:} Zoom-in to the central $10 {h^{-1}{\rm Mpc}}$
  region. The mass of each particle in this re-simulation was 64 times
  smaller than in the simulation shown in the left panel. A large halo
  just at the center has a mass $\sim 10^{12}M_\odot$. Its environment
  is similar to the invironment of our galaxy. Note that a large
  filamentary structure, which goes from the left top corner to the bottom
  right corner is composed of numerous small filaments. A variety of
  dark matter haloes is found along and at the intersections of different
  filaments. }
\label{fig:slice160}
\end{figure}

In Fig. \ref{fig:slice160}, left panel,  we show an
example of the density field in one of the simulations. The density
distribution is remarkably complex.  There are large quasi-spherical
under-dense regions of different sizes.  The dense regions show two
types of structures. When the density is very large the structures are
nearly spherical (typical axial ratios are 1:1.5 - 1:2). Those are
called halos.  There are numerous filaments, which have lower
density. The filaments contain chains of halos with the largest halos
placed at intersections of filaments.

To study in more detail the properties of objects, we also performed
high mass resolution simulations of selected regions. The right panel
of Fig. \ref{fig:slice160} shows an example of such a simulation where we
selected a sphere of approximately mean density close to the center of
the simulation box shown in Fig. \ref{fig:slice160}, left panel. The mass
resolution in the low-resolution simulation is $2.6 \times 10^8
h^{-1}{\rm {M_{\odot}}}$: the whole $160{h^{-1}{\rm Mpc}}$ box was simulated with $1024^3$
particles. In the re-simulated region of radius $\sim 15{h^{-1}{\rm Mpc}}$ the
mass resolution is $4.0 \times 10^6 {h^{-1}{\rm {M_{\odot}}}}$. 
Therefore, objects
similar to the Local Group are resolved with almost 1 million
particles. One can see a clear difference between the left and right
panels in Fig. \ref{fig:slice160}: in the small region, an 
environment typical for our Galaxy,  there are many tiny filaments
pointing in the same direction, which is also the direction of the
large-scale velocity field. Hundreds of small halos 
($10^8 {h^{-1}{\rm {M_{\odot}}}}$ to
$10^9 {h^{-1}{\rm {M_{\odot}}}}$) are strung together along these filaments.

The identification of haloes is always a challenge. We have developed
two algorithms: the hierarchical friends-of-friends (HFOF) and the
bound density maxima (BDM) algorithms \cite{klypin:1999}.  Both were
parallelized using MPI (FOF) or OpenMP (BDM). They are complementary
and find essentially the same haloes. Thus, we believe that the
algorithms are stable and capable of identifying all dark matter
haloes in the simulations.  The advantage of the HFOF algorithm is
that it can handle haloes of arbitrary shape at arbitrary
over-density, not just spherical haloes.  The advantage of the BDM
algorithm is that it describes the physical properties of the haloes
better by identifying and removing unbound particles. This is 
particularly important for finding sub-halos.

\begin{figure}[t]
\sidecaption[t]
\includegraphics[width=0.5\textwidth]{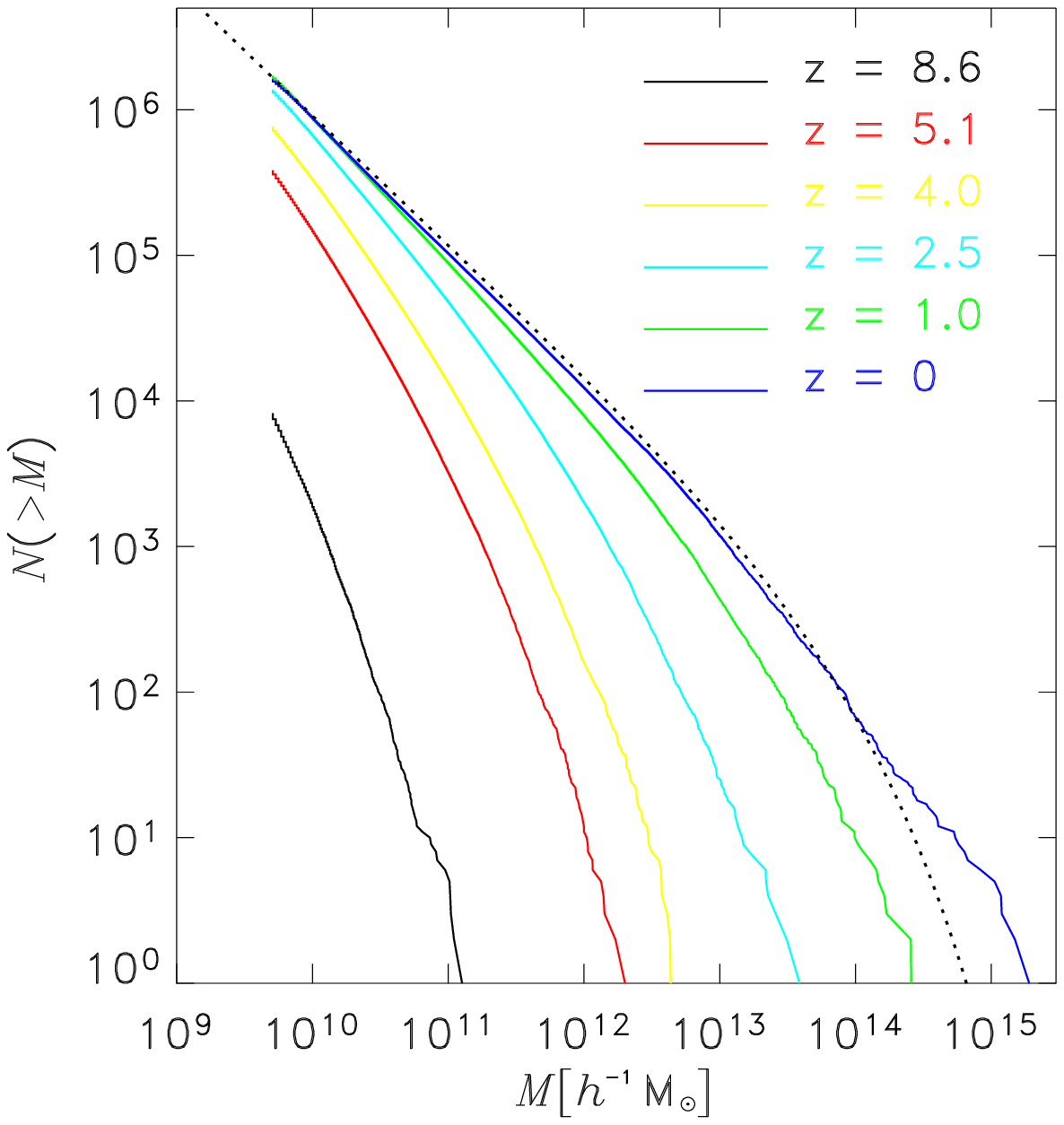}
\includegraphics[width=0.5\textwidth]{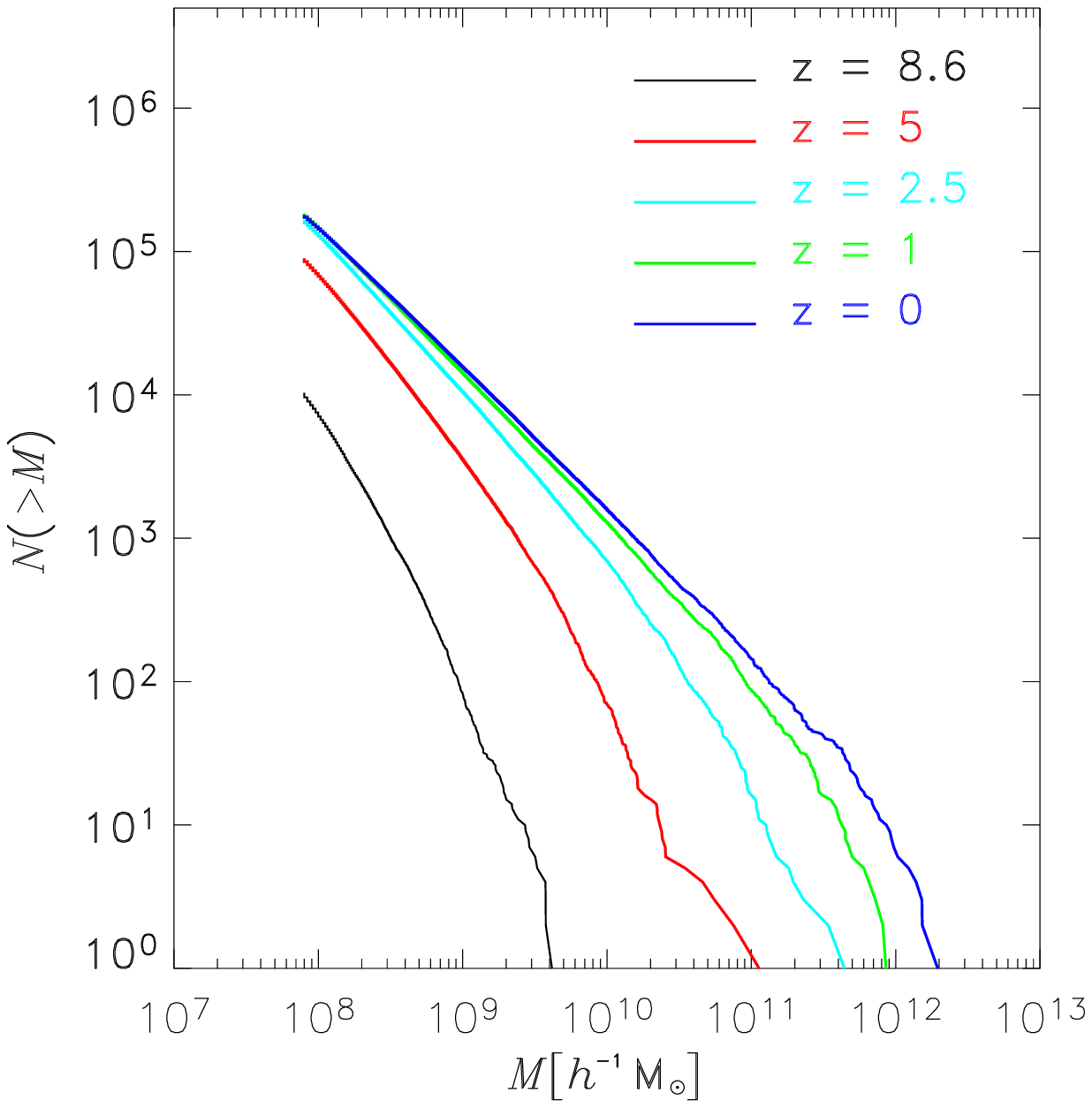}
\caption{{\it Left:} The mass function of FOF-halos detected at
  different redshifts in the simulation shown in the left panel of
  Fig. \ref{fig:slice160}. The dotted curve is analytical
  approximation \cite{Tinker2008}.  {\it Right:} The mass
  function in the high resolution area shown in the right panel of
  Fig. \ref{fig:slice160} .}
\label{fig:mass_function}
\end{figure}

In Fig.  \ref{fig:mass_function} we show the mass function of halos
detected with the friends-of-friends algorithm. We have identified in
the full box already at redshift $z=8.6$ more than 8000 halos and at
redshift $z=0$ more than 1.6 million halos (left panel). Due to the better mass
resolution in the re-simulation region (a sphere of $15{h^{-1}{\rm Mpc}}$ at
redshift $z=0$) we can identify 64 times less massive halos. We found
at redshift $z=8.6$ more than 10,000 halos, and at redshift $z=0$
almost 200,000 halos (right panel).

In \cite{prada:2006} we used a high resolution re-simulation of a
filament with 150,000,000 particles as well as simulations of a full
boxes of 80 and $120 {h^{-1}{\rm Mpc}}$ size with $512^3$ particles to study
isolated halos. With a mass resolutions of 
$4.9 \times 10^6 {h^{-1}{\rm {M_{\odot}}}}$, 
$3.2 \times 10^8 {h^{-1}{\rm {M_{\odot}}}}$ or 
$1.1 \times 10^9 {h^{-1}{\rm {M_{\odot}}}}$ density
profiles of collapsed galaxy-size dark matter halos with masses
$10^{11}-5\cdot 10^{12}{\rm {M_{\odot}}}$ can be measured very accurately. We
found that isolated halos in this mass range extend well beyond the
formal virial radius $R_{\rm vir}$ exhibiting all properties of
virialized objects up to 2--3$R_{\rm vir}$: relatively smooth density
profiles and no systematic infall velocities. Contrary to more massive
halos, the dark matter halos in this mass range do not grow through a
steady accretion of satellites. For larger radii we combine the
statistics of the initial fluctuations with the spherical collapse
model to obtain predictions for the mean and most probable density
profiles.  The model gives excellent results beyond 2-3 formal virial
radii.

Based on a simulation of a $150 {h^{-1}{\rm Mpc}}$ box we  studied the
efficiency of different approaches to interloper treatment in
dynamical modeling of galaxy clusters \cite{wojtak:2007}.  Taking
advantage of the full 3D information available from the simulation, we
selected samples of interlopers defined with different criteria to
assess the efficiency of different interloper removal schemes.  We
found that the direct methods exclude on average 60-70 percent of
unbound particles producing a sample with contamination as low as 2-4
percent. Using indirect approaches, which are applied to the data
stacked from many objects, we reproduced the properties of composite
clusters and estimated the probability of finding an interloper as a
function of distance from the object center. We used mock catalogs
extracted from the same simulation to test a new method with which we
studied the mass distribution in six nearby ($z<0.06$) relaxed Abell
clusters of galaxies \cite{lokas:2006b}. Based on this cosmological
$N$-body simulation we are able to interpret the complex velocity
distribution of galaxies in galaxy cluster Abell 1689 \cite{lokas:2006a}.

\begin{acknowledgement} The computer simulations described here have
been performed at LRZ Munich, BSC Barcelona and NAS Ames.  We
acknowledge support of NSF and NASA grants to NMSU and DAAD support of
our collaboration.  We thank A. Kravtsov (University of Chicago), 
G. Yepes (UAM, Madrid), A. Khalatyan
(AIP, Potsdam), and Y. Hoffman (HU, Jerusalem) for helpful
discussions.

\end{acknowledgement}
\end{document}